\documentclass[square]{ws-procs975x65}

\begin{document}

\title{STOCHASTIC STABILITY AND THE SPIN GLASS PHASE.\\
THE STATE OF ART FOR MEAN FIELD AND  
FINITE DIMENSIONAL MODELS}

\author{P. CONTUCCI}

\address{Department of Mathematics, Alma Mater Studiorum\\ 
University of Bologna, Italy\\
E-mail:
pierluigi.contucci@unibo.it\\
http://www.dm.unibo.it/$\sim$contucci}

\begin{abstract}
Some invariances under perturbations of the spin glass phase are
introduced, their proofs outlined and their consequences illustrated 
as factorization rules for the overlap distribution. A comparison between
the state of the art for mean field and finite dimensional models
is shortly discussed.
\end{abstract}


\bodymatter
\vspace{2cm}
Factorisation laws for observable quantities are very important both from theoretical 
and experimental perspective in statistical mechanics and quantum field theory. Their 
rigorous proof is then a crucial step toward the understanding of a theory and its specific models. 
Examples are the Wick law in Euclidian quantum field theory, the multiplicative factorisation
of the Curie Weiss theory of ferromagnetism, and more recently, the Parisi theory of 
the spin glass phase. This last case presents a factorisation structure that has been one of 
the most intriguing features of theoretical physics in the last decades and has found applications in many
areas as diverse as computer sciences, biology and economics by means of new ideas and methods
in probability, machine learning, optimisation etc. 

In this short note we give a brief account of its nature and its origins introducing, via a 
mathematically elementary method based on sound physical ideas, both its form and a guide to its proof. 

We consider configurations of $N$ Ising spins
\begin{equation}
\sigma  =  \{\sigma_i\}, \quad \tau  =  \{\tau_i\}, \quad ...
\end{equation}
and introduce a centered Gaussian Hamiltonian $H_N(\sigma)$ defined by the covariance 
\begin{equation}
{\rm Av}(H_N(\sigma)H_N(\tau))  =  N c_N(\sigma,\tau)  .
\end{equation}
For all $\beta>0$ we are interested in the random probability measure
\begin{equation}
p_N(\sigma) =  \frac{e^{-\beta H_N(\sigma)}}{\sum_\sigma e^{-\beta H_N(\sigma)}}
\end{equation}
Important examples are:\\
the Sherrington-Kirkpatrick model
\begin{equation}
c_N(\sigma,\tau) =  \left( \frac{1}{N}\sum_{i=1}^{N}\sigma_i\tau_i\right)^2  
\end{equation}
and the Edwards-Anderson model 
\begin{equation}
c_N(\sigma,\tau) =  \frac{1}{dN}\sum_{|i-j|=1}\sigma_i\sigma_j\tau_i\tau_j  .
\end{equation}
Quantities of interest include:
\begin{itemlist}
\item the pressure
\begin{equation}
P_N(\beta) =  {\rm Av} \log \sum_\sigma e^{-\beta H_N(\sigma)}
\end{equation}
\item the moments
\begin{equation}
{\rm Av}\left(\frac{\sum_{\sigma,\tau}c_N(\sigma,\tau)e^{-\beta[H_N(\sigma)+H_N(\tau)]}}{\sum_{\sigma,\tau}e^{-\beta[H_N(\sigma)+H_N(\tau)]}} \right)  = 
\end{equation}
\begin{equation}\nonumber
=  <c>_N  :=  \int c p_N(c)dc  \; ,
\end{equation}
\begin{equation}
{\rm Av}\left(\frac{\sum_{\sigma,\tau,\gamma}c_N(\sigma,\tau)c_N(\tau, \gamma)e^{-\beta[H_N(\sigma)+H_N(\tau)+H_N(\gamma)]}}{\sum_{\sigma,\tau,\gamma}e^{-\beta[H_N(\sigma)+H_N(\tau)+H_N(\gamma)]}} \right) =  <c_{12}c_{23}>_N  =
\end{equation}
\begin{equation}\nonumber
:=  \int c_{12}c_{23}p^{(12),(23)}_N(c_{12},c_{23})  ,
\end{equation}
etc. and, especially, the
\item \underline{joint overlap distribution}
\begin{equation}
p^{(12),(23),...,(kl),...}_N(c_{12},c_{23},...,c_{kl},...)  .
\end{equation}
\end{itemlist}
The Parisi theory for spin glasses (also known as Replica Symmetry Breaking theory), 
is based on the following {\it ``factorization''} \underline{assumptions} (see \cite{mpv})
in the thermodynamic limit:
\begin{itemize}
\item Replica Equivalence, e.g.:
\begin{equation}
p^{(12),(23)}(c_{12},c_{23})  =  \frac{1}{2}p(c_{12})\delta(c_{12}-c_{23})+\frac{1}{2}p(c_{12})p(c_{23})
\end{equation}
\begin{equation}
p^{(12),(34)}(c_{12},c_{34})  =  \frac{1}{3}p(c_{12})\delta(c_{12}-c_{34})+\frac{2}{3}p(c_{12})p(c_{34})
\end{equation}
\item Ultrametricity, e.g.:
\begin{equation}
p^{(12),(23),(31)}(c_{12},c_{23},c_{31}) =
\end{equation}
\begin{equation}\nonumber 
\delta(c_{12}-c_{23}) \delta(c_{23} - c_{31}) p(c_{12})\int_0^{c_{12}}dcp(c)
\end{equation}
\begin{equation}\nonumber 
+ \theta(c_{12}-c_{23})\delta(c_{23} - c_{31})p(c_{12})p(c_{23})
\end{equation}
\begin{equation}\nonumber 
+ \; {\rm 2} \; {\rm cyclic} \; {\rm permutations}
\end{equation}
\end{itemize}

Replica Equivalence and Ultrametricity allows to reconstruct the 
$p^{(\{kl\})}(\{c_{kl}\})$ starting from $p(q)$. This clearly means that 
we have a complete factorization and we are in presence of a typical
mean field picture. Similar examples were the validity of the Wick rule
for fields expectation in Euclidean quantum field theory, or the 
Boltzmann-Gibbs complete factorization for the Curie Weiss model.

We summarise here the results (see \cite{cgg}) on a rigorous version of the
linear response theory applied to the spin glass phase. The stability method in Statistical Mechanics works by identifying 
a small (yet non-trivial) deformation of the system, prove that in the 
large volume limit the perturbation vanishes and, by means of the linear response theory,
compute the relations among observable quantities. This method leads to interesting
consequences and applications because it reduce the apriori degrees of freedom of a theory.
For instance it shows that ferromagnetic mean field models have a magnetisation which is
a full order parameter for the theory.

More specifically in classical models one starts, for smooth bounded functions $f$ of spin configurations, 
from the counting measure
\begin{equation}
\mu_N(f)  =  \frac{1}{2^N}\sum_{\sigma}f(\sigma)  ,
\end{equation}
and defines the equilibrium state
\begin{equation}
\omega_{\beta,N}(f)  =  \frac{\mu_N(fe^{-\beta H_N})}{\mu_N(e^{-\beta H_N})}  ;
\end{equation}
By considering the Hamiltonian per particle
\begin{equation}
h_N(\sigma)=\frac{H_N(\sigma)}{N}
\end{equation}
the classical perturbed state is defined by
\begin{equation}
\omega_{\beta,N}^{(\lambda)}(f)  =  \frac{\omega_{\beta,N}(fe^{-\lambda h_N})}{\omega_{\beta,N}(e^{-\lambda h_N})}  .
\end{equation}
Since the perturbation amounts to a small change in the temperature
\begin{equation}
\omega_{\beta,N}^{(\lambda)}(f)  =  \omega_{\beta +\frac{\lambda}{N},N}(f)
\end{equation}
one has that, a part for isolated singularity points, in the thermodynamic limit
\begin{equation}
\frac{d \omega_{\beta,N}^{(\lambda)}(f)}{d\lambda}  \rightarrow  0  .
\end{equation}
One may easily show that the previous property implies that 
for the Curie-Weiss model in zero magnetic field
\begin{equation}
\omega_{\beta}(\sigma_1\sigma_2\sigma_3\sigma_4)  =  \omega_{\beta}(\sigma_1\sigma_2)^2  \; .
\end{equation}
The previous approach lead to the concept of {\it Stochastic Stability} when applied,
suitably adapted, to the spin glass phase. Consider, for smooth bounded function $f$ of $n$ spin
configurations, the quenched equilibrium state
\begin{equation}
< f >_{\beta,N}  =  
{\rm Av}\left(\frac{\sum_\sigma f(\sigma)e^{-\beta H_N}}{\sum_\sigma e^{-\beta H_N}}\right)
\end{equation}
define the deformation as:
\begin{equation}
< f >^{(\lambda)}_{\beta,N}  =  \frac{<fe^{-\lambda h_N}>}{<e^{-\lambda h_N}>}  .
\end{equation}
We observe that the previous deformation is {\bf not} a simple temperature shift. In fact:
\begin{equation}
< f >^{(\lambda)}_{\beta,N}  =  
\frac{{\rm Av}\left(\frac{\sum_\sigma f(\sigma)e^{-(\beta+\lambda/N)H_N}}{\sum_\sigma e^{-\beta H_N}}\right)}
{{\rm Av}\left(\frac{\sum_\sigma e^{-(\beta+\lambda/N)H_N}}{\sum_\sigma e^{-\beta H_N}}\right)}  ,
\end{equation}
nevertheless the system {\bf is still} stable under it.
We can state our main result as follows. The spin glass quenched equilibrium state is stable 
with respect to the deformation defined above in the sense that, a part for isolated singularity points, 
in the thermodynamic limit
\begin{equation}
\frac{d}{d\lambda} < f >^{(\lambda)}_{\beta,N}  \to  0  ;
\end{equation}
moreover the previous stability property implies (by use of the integration by parts techinque) 
that the following set of identities (Ghirlanda-Guerra) holds: 
\begin{equation}
< f c_{1,n+1}>_{\beta,N}   =  \frac{1}{n}< f >_{\beta,N}< c >_{\beta,N} + 
\end{equation}
\begin{equation}\nonumber
+ \frac{1}{n}\sum_{j=2}^{n}< fc_{1,j} >_{\beta,N}  \; ,
\end{equation}
where the term $c_{1,n+1}$ is the overlap between a spin configuration of the set $\{1,2,...,n\}$ and and external
one that we enumerate as $(n+1)$-nth, and $c_{1,j}$ is the overlap between two generic spin configurations among
the $n$'s. 

The proof ideas can be easily summarized by the study of three quantities
and their differences which encode the fluctuation properties of the spin
glass system:
\begin{equation}
{\rm Av}\left[\omega(H_N^2)\right]  \; ,
\end{equation}
\begin{equation}
{\rm Av}\left[\omega(H_N)^2\right]  \; ,
\end{equation}
\begin{equation}
{\rm Av}\left[\omega(H_N)\right]^2  \; .
\end{equation}
The result is obtained by two bounds:
\begin{itemlist}
\item bound on averaged thermal fluctuations
\begin{equation}
{\rm Av}\left[\omega(H_N^2)-\omega(H_N)^2\right] \le c_1 N
\end{equation}
obtained by stochastic Stochastic Stability method (see \cite{ac}) by showing that
the addition of an independent term of order one to the Hamiltonian 
is equivalent to a small change in temperature of the entire system:
\begin{equation}
\beta H_N(\sigma) \to \beta H_N(\sigma) + \sqrt{\frac{\lambda}{N}}{\tilde H}_N(\sigma)
\end{equation}
\begin{equation}
\beta \to \sqrt{\beta^2 +\frac{\lambda}{N}}
\end{equation}
\item bound on disorder fluctuations
\begin{equation}
U=\omega(H_N)
\end{equation}
\begin{equation}
{\rm Av}(U^2)-{\rm Av}(U)^2 \le c_2 N  ,
\end{equation}
which is the self averaging of internal energy and can be proved from self averaging
of the free energy (with martingale methods or concentration of measures).
\end{itemlist}

A recent achievement (see \cite{p}) shows that
the validity of the previous identities extended to the moments of all orders, i.e.
their validity in distribution, is compatible only with an ultrametric distribution of 
the overlaps. Since the validity of the distributional identities was previously shown
by Talagrand for the mean field models, Panchenko's result proves the ultrametricity
result for those models. 

The next challenge is to work with models in finite dimensions (especially $d=3$) 
who are more directly related to physics. For those models a proof of the distributional
factorisation laws is in progress \cite{cms}. One has to stress nevertheless
that its achievement wouldn't end the discussion about the triviality issue of the 
Edwards-Anderson model because it leaves still open the nature of the single
overlap distribution. The rigorous proof for its non-triviality, available so far,
works only for the general mean field case.\\\\

{\bf Acknowledgments}. This work was presented at the ICMP12 in Aalborg. It is a pleasure
to thank the organisers for both the invitation and the opportunity to give the talk and to 
write this contribution.

\end{document}